\documentclass[prb,aps,twocolumn,amsmath,amssymb,superscriptaddress]{revtex4-1}

 \usepackage{amsmath}
\usepackage{tabularx}
 \usepackage{graphicx}
\usepackage{multirow}
 \usepackage{SIunits}
 \usepackage{ulem}
\usepackage{braket}
 \usepackage{color}
\usepackage{amssymb}
\usepackage{dcolumn}
\usepackage{bm}
\newcolumntype{P}[1]{>{\centering\arraybackslash}p{#1}}

\begin{document}

\title{Combined atomic force microscopy and photoluminescence imaging to select single InAs/GaAs quantum dots for quantum photonic devices}

\date{\today}
\author{Luca Sapienza}
\email{l.sapienza@soton.ac.uk}
\affiliation{Department of Physics and Astronomy, University of Southampton, Southampton SO17 1BJ, UK}
\affiliation{Center for Nanoscale Science and Technology, National Institute of Standards and Technology, Gaithersburg, MD 20899, U.S.A.}

\author{Jin Liu}
\affiliation{Center for Nanoscale Science and Technology, National Institute of Standards and Technology, Gaithersburg, MD 20899, U.S.A.}
\affiliation{Maryland NanoCenter, University of Maryland, College Park, MD 20742, USA}

\author{Jin Dong Song}
\affiliation{Center for Opto-Electronic Materials and Devices Research, Korea Institute of Science and Technology, Seoul 136-791, South Korea}

\author{Stefan F$\ddot{\text{a}}$lt}
\affiliation{Solid State Physics Laboratory, Swiss Federal Institute of Technology, 8093 Zurich, Switzerland}

\author{Werner Wegscheider}
\affiliation{Solid State Physics Laboratory, Swiss Federal Institute of Technology, 8093 Zurich, Switzerland}

\author{Antonio Badolato}
\affiliation{
Department of Physics and Max Planck Centre for Extreme and Quantum Photonics, University of Ottawa, Ottawa, Ontario K1N 6N5, Canada}

\author{Kartik Srinivasan}
\email{kartik.srinivasan@nist.gov}
\affiliation{Center for Nanoscale Science and Technology, National Institute of Standards and Technology, Gaithersburg, MD 20899, U.S.A.}

\begin{abstract}

We report on a combined photoluminescence imaging and atomic force microscopy study of single, isolated self-assembled InAs quantum dots (density $<0.01$ $\mu$m$^{-2}$) capped by a 95\,nm GaAs layer, and emitting around 950\,nm. By combining optical and scanning probe characterization techniques, we determine the position of single quantum dots with respect to comparatively large ($100$~nm to $1000$~nm in-plane dimension) topographic features. We find that quantum dots often appear ($\gtrsim~25~\%$ of the time) in the vicinity of these features, but generally do not exhibit significant differences in their non-resonantly pumped emission spectra in comparison to quantum dots appearing in defect-free regions. This behavior is observed across multiple wafers produced in different growth chambers. Our characterization approach is relevant to applications in which single quantum dots are embedded within nanofabricated photonic devices, where such large surface features can affect the interaction with confined optical fields and the quality of the single-photon emission.  In particular, we anticipate using this approach to screen quantum dots not only based on their optical properties, but also their surrounding surface topographies.
\end{abstract}

 \maketitle

\section{Introduction}

Single self-assembled InAs/GaAs quantum dots (QDs) grown by molecular beam epitaxy (MBE) \cite{Gerald, Petroff} are one of the most promising solid-state emitters for quantum technologies, due to their potential high stability and emission efficiency, easy on-chip integration, and coherence of the single-photon emission \cite{Glenn}. To preserve such characteristics, QDs have to be capped by larger bandgap semiconductors (e.g., GaAs), with layer thicknesses typically exceeding  50\,nm \cite{Antonio_cap}. To study the relation between material structure and optical properties of QDs, several techniques have been employed, including scanning probe microscopy \cite{AFM_QDs, AFM_QDs2, AFM_QDs3, STM} of uncapped or partially capped QDs, and transmission electron microscopy of capped QDs\cite{TEM_QD, TEM_QD2}. Advances in crystal growth have reached high structural control and material purity, allowing QDs to be successfully employed as gain media in lasers \cite{QD_lasers} and as single artificial atoms in cavity quantum electrodynamics\cite{Glenn, JMG}.

However, photonic devices (such as microcavities) that require a single emitter to be in a certain position and to emit at a specific wavelength are still very challenging to implement with high yield with these QDs, because the Stranski-Krastanov nucleation process at the origin of QD growth produces a random spatial positioning of the QDs across the wafer and inhomogeneous spectral broadening of the QD ensemble exciton emission. To achieve accurate positioning of single QDs within nanophotonic devices, two classes of techniques have been developed, one based on changes to the surface morphology in correspondence to the buried emitters, and the other based on their light emission. The former includes atomic force microscope (AFM) mapping~\cite{Hennessy} and scanning electron microscopy~\cite{AB_SEM, Iwamoto} to detect surface deformations due to strain propagation from the buried QD, while the latter includes scanning confocal photoluminescence microscopy~\cite{Lee,Thon_APL_09,Dousse}, scanning cathodoluminescence~\cite{3D_lenses}, and photoluminescence imaging~\cite{Kojima, NComm}. In this work, we combine techniques from each of these two classes in order to better understand the extent to which the optical performance of single QD nanophotonic devices might be influenced by the morphology of the crystal structure surrounding the QD.

\begin{figure*}[t]
  \includegraphics[width=0.85\linewidth]{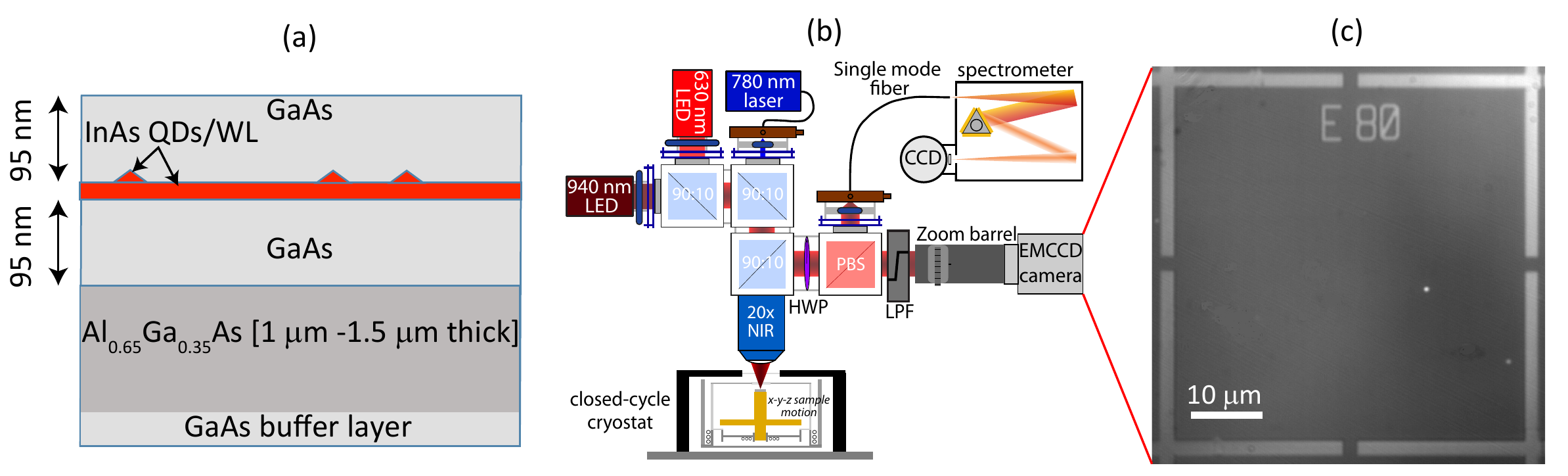}
   \caption{(a) Schematic of the sample under study (not to scale), comprising a single layer of InAs quantum dots (red triangles), grown on an InAs wetting layer (WL) between two 95~nm thick layers of GaAs, and situated on top of a 1~$\mu$m or 1.5~$\mu$m thick Al$_{0.65}$Ga$_{0.35}$As layer on a GaAs buffer layer followed by a GaAs substrate. (b) Schematic of the photoluminescence setup. An infrared light emitting diode (LED, emission centered at 940~nm) is used for illumination of the sample while either a 630~nm red LED or a 780~nm laser is used for excitation of the quantum dots (QDs), depending on whether excitation over a broad area (LED) or of individual QDs (laser) is required. Samples are placed within a cryostat on an x-y-z positioner. Imaging is done by directing the emitted and reflected light into an Electron Multiplied CCD (EMCCD) camera, while spectroscopy is performed by collecting emission into a single-mode fiber and sending it to a grating spectrometer. (c) EMCCD image of the photoluminescence from two QDs and reflected light by the alignment marks (metallic crosses), acquired by illuminating the sample simultaneously with both the red and near-infrared LEDs, at a temperature of 4~K.}
   \label{fig1}
\end{figure*}

The presence of surface features in QD epitaxy is not surprising, given the number of interfaces present in a layer structure such as that shown in Fig.~\ref{fig1}(a), and the potential for defects to form at interfaces and influence (through strain) the growth of subsequent layers. Here, the InAs QD layer is sandwiched between two GaAs layers, which are grown on a thick Al$_{0.65}$Ga$_{0.35}$As layer (used as a sacrificial layer in device fabrication) that is grown on a GaAs substrate. The coherent deformation of the crystal caused by the InAs/GaAs lattice mismatch produced when growing the QD layer is intrinsic and appears on the surface as a shallow island with sub-micrometer diameter. In contrast, defects formed well below the InAs layer, close to the GaAs substrate/epilayer interface, are usually buried by thick lattice-matched buffer layers (e.g., the Al$_{0.65}$Ga$_{0.35}$As layer) and can recover the crystal coherence, appearing at the surface as large convex oval defects \cite{defects}.  Because nucleation of InAs is energetically favorable close to crystal steps, QDs tend to decorate the edge of the oval defects. The coherent deformations of the crystal surrounding a QD and propagating up to the surface can strongly affect the QD emission properties without necessarily affecting the optical quality (e.g, as judged by the emission linewidth or number of emitting states). This aspect is often neglected in experiments relying on random choice of the target QD, but may be important when considering the behavior of the QD within a surrounding photonic structure. In some cases, such crystal deformation can have positive benefit.  For example, in Ref.~\,\onlinecite{QD_oval}, the authors attributed an increased emission intensity to the presence of an unclassified oval defect close to the QD, acting  in a similar way as a solid immersion lens and thus increasing the extraction efficiency of the single photons emitted by the QD~\cite{SIL}.  On the other hand, crystal defects may degrade the performance of nanofabricated photonic structures, such as photonic crystals, which are used to enhance radiative rates and extraction efficiency, but whose characteristic lengths are at the sub-micrometer scale, and are thus comparable to the surface defect sizes.

Here, we present a study of the surface morphology of GaAs samples containing single QDs and investigate the correlation between surface features and emitter locations. This is made possible by the implementation of a photoluminescence imaging technique that we have recently developed, based on a double light-emitting diode (LED) illumination scheme (see Fig.\,1(b)). By using a red LED to excite the QD emission and near-infrared LED to illuminate alignment marks deposited on the sample's surface, we are able to locate single QDs with uncertainties below 30\,nm \cite{NComm}. Such a technique allows us to optically characterize the emitters and find their positions (see Fig.\,1(c)), and then to investigate the nearby surface morphology of the sample by means of AFM. By combining photoluminescence and AFM techniques, we are able to correlate the position of the QDs with respect to the surface features observed.

\section{Experimental Results}

We have investigated four QD samples (labeled as samples I-IV) presenting low QD densities (about 1 to 10 QDs per 1000 $\mu$m$^{2}$), all with the same nominal structure (see Fig.\,1(a)) and emitting at wavelengths between 900~nm and 1000\,nm, but grown in different MBE chambers. We focus on such ultra-low QD density materials due to their specific relevance to quantum photonic experiments in which, for example, the interaction between only one QD and a confined optical mode must be ensured. Previous experiments operating in this regime include studies of strong coupling cavity QED \cite{Winger} and triggered single-photon generation \cite{NComm}.

\begin{figure*}
  \includegraphics[width=0.7\linewidth]{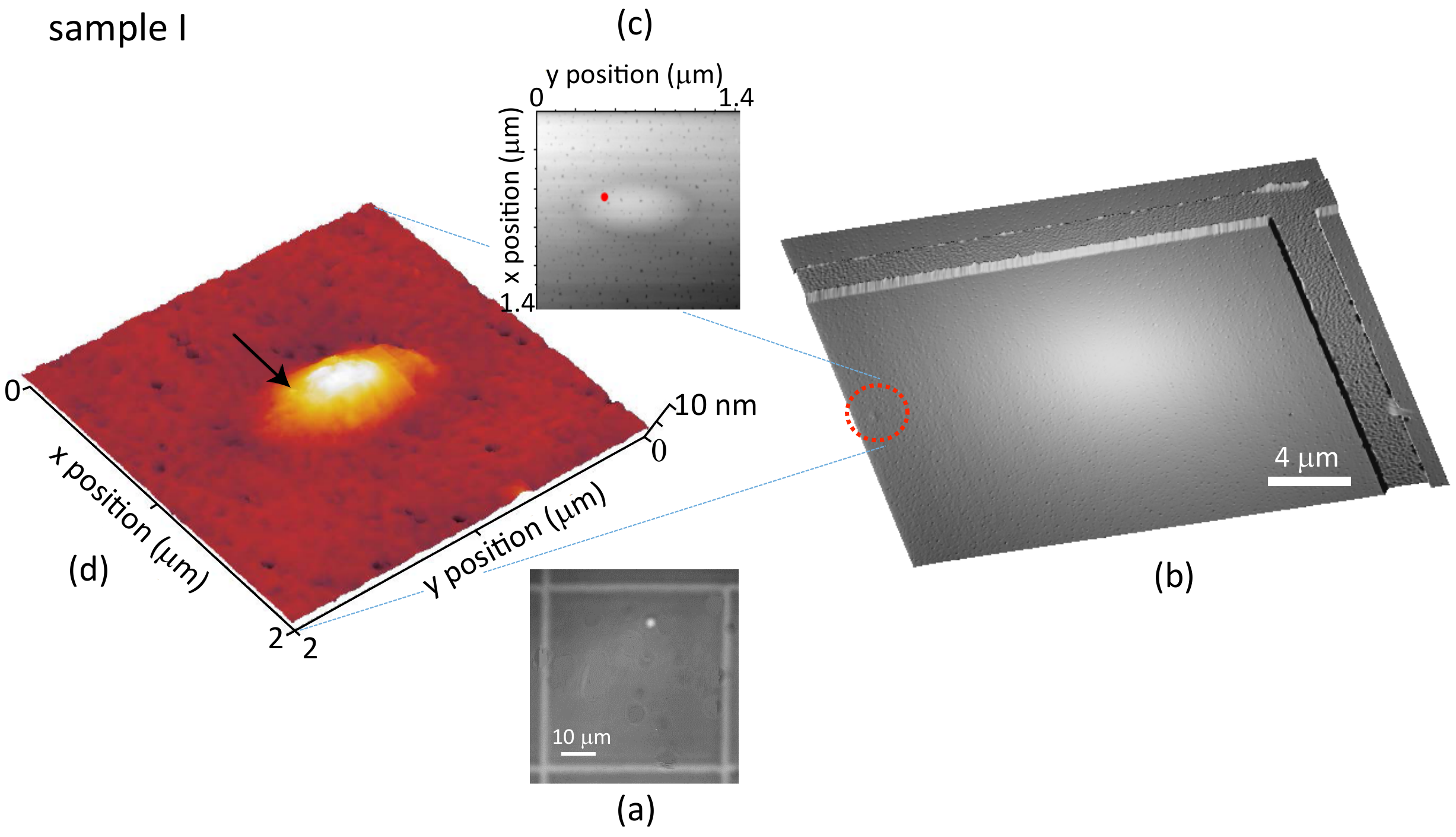}
   \caption{(a) EMCCD image of the photoluminescence from a single QD and reflected light by the alignment marks (metallic crosses), acquired by illuminating the sample simultaneously with both the red and near-infrared LEDs, at a temperature of 4~K. (b) Atomic force microscope image of the area between two alignment marks (top right corner of panel (a)). (c,d) Atomic force microscope images of a surface oval defect on which the position of the QD, measured from the photoluminescence image (red symbol in panel (c) and arrow in panel (d)), is shown. The one standard deviation uncertainty in the position of the QD is estimated to be 17~nm (see main text). [Sample I]}
   \label{fig2}
\end{figure*}

Samples I, III, and IV were grown by MBE at ETHZ. We used cracked As tetramer and a GaAs growth rate of 2.2~$\text{\AA}$/s. The Ga cell featured a dual filament and was operated with cold tip (i.e., the ultra-high electron mobility configuration). The substrate temperature during the growth of the 160~nm GaAs buffer layer was kept above 600~$^{\circ}$C and it was lowered to 580~$^{\circ}$C during the growth of the 190~nm GaAs device layer. The 1~$\mu$m thick AlGaAs sacrificial layer was either grown by digital alloy (samples I and III) or by co-growth, changing the Ga cell temperature (sample IV). The AlGaAs growth rate was always below 2.8~$\text{\AA}$/s. The QDs were grown at 520~$^{\circ}$C and the rotation of the substrate was stopped to create a gradient in the QD density over the wafer. Sample II was grown by MBE at KIST. We used cracked As tetramer and a GaAs growth rate of 1.4~$\text{\AA}$/s (Ga cell 1). The two Ga cells featured a dual filament and were operated with hot tip. The substrate temperature during the growth of the GaAs layers (200~nm thick buffer layer and 190~nm thick device layer) and 1.5~$\mu$m thick AlGaAs layer was kept at 590~$^{\circ}$C. The growth rates for the AlGaAs sacrificial layer, created through co-growth, were 0.68~$\text{\AA}$/s for GaAs (Ga cell 2) and 1.47~$\text{\AA}$/s for AlAs. The QDs were grown at the substrate temperature of 510~$^{\circ}$C and the rotation of the substrate was stopped to create a gradient in the QD density over the wafer.

After locating the emission of single QDs using the photoluminescence imaging method (see Fig.\,2(a)), we investigate the surface morphology in the nearby area via AFM (see Fig.\,2(b)). The AFM images that we obtain show different surface features and an example is the oval defect shown in Fig.\,2c and 2d (sample I). Such defects are attributed to GaAs droplets that can be formed during MBE growth \cite{QD_oval, Joanna} and are assumed to appear because of Ga cell spitting \cite{oval}. The presence of a variety of surface defects in MBE grown GaAs samples has been extensively reported (see, for instance, Refs.\,\onlinecite{surf_defects, surf_defects2, surf_defects3, surf_defects4}) and attributed mostly to the operation and geometry of the Ga cell.
In our technique, we first acquire a photoluminescence image of the sample so as to determine the location of the QD emitting dipole with respect to the middle point of the two nearest alignment marks (Fig.\,2(a)). Then, the sample's surface, in correspondence to the area where the QD had been optically located, is mapped by using an AFM equipped with a tetrahedral, point-terminated, silicon cantilever with a tip radius of 7\,nm, spring constant of 26\,N/m and resonance frequency of 300\,kHz (nominal values), in tapping mode (Fig.\,2(b)). The sample is aligned in such a way that the scanning tip direction is orthogonal to one of the alignment marks. Two orthogonal scans are then collected (tip scanning at 0$^{\circ}$ and 90$^{\circ}$ angles) in order to be able to image the edges of the two orthogonal alignment marks with high accuracy. When a surface feature is observed, the position of the QD (distance from the center of the alignment marks) extracted from the optical images is marked on the AFM image (Fig.\,2(c)). If needed, a zoomed-in scan is collected and the position of the surface feature with respect to the alignment marks is recovered by taking into account the scanning offset introduced in the process (Fig.\,2(d)). We estimate the uncertainty in specifying the position of the emitting dipole in the AFM scan (Fig.~\ref{fig2}(c)) based on the combined one standard deviation uncertainty of locating features in the optical positioning technique and the AFM scan.  For the optical positioning technique, the location of the emitting dipole with respect to the center of the nearest alignment mark is determined from Gaussian fits to line cuts through the optical image~\cite{NComm}.  For the AFM data, the alignment mark center is determined from Gaussian fits to the two peaks found in the first derivative of a line cut through the AFM image. For the data shown in Fig.~\ref{fig2}(c), the imaging and AFM uncertainties are 11~nm and 13~nm, respectively, giving a combined uncertainty of 17~nm.

\begin{figure*}[t]
\includegraphics[width=\linewidth]{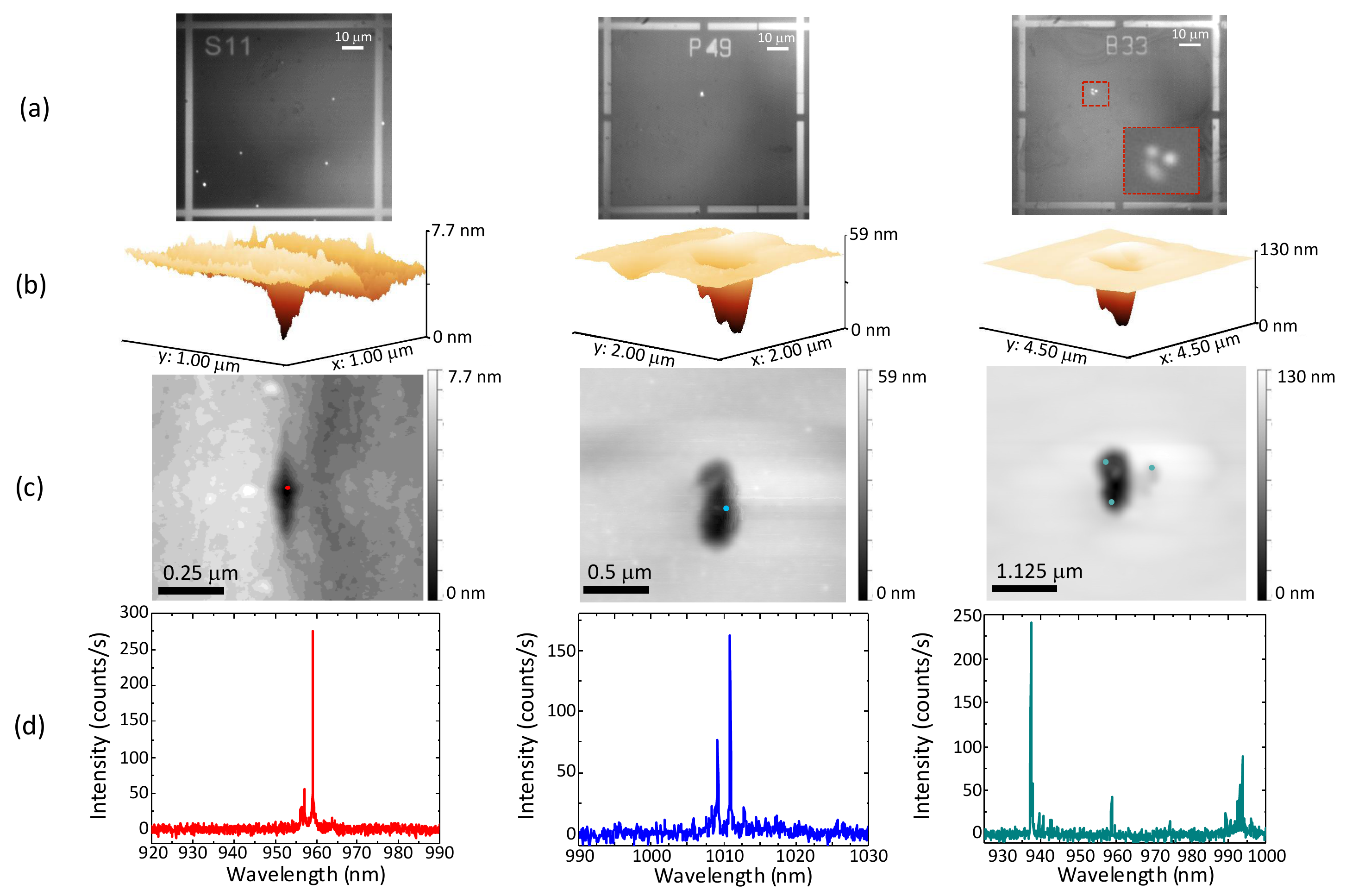}
   \caption{(a) EMCCD images of the photoluminescence from QDs and reflected light by the alignment marks (metallic crosses), acquired by illuminating three different fields of two different samples (sample II in the first column and sample III in the second and third columns) simultaneously with both the red and near-infrared LEDs, at a temperature of 4~K. The inset in the third image represents a zoom-in of the area marked by the dashed red lines in the QD photoluminescence image. (b) Three-dimensional atomic force microscope images of surface features located in proximity to the QD's emitting dipole positions, obtained from the photoluminescence images shown in panel (a) (the images are rotated 50$^{\circ}$ counterclockwise with respect to the images in panel (c)). (c)  Two-dimensional atomic force microscope images of the surface features shown in panel (b). The colored dots represent the QD locations, as extracted from the images in panel (a). The one standard deviation uncertainties in the QD location are 47~nm (Sample II, left panel), 31~nm (Sample III, center panel), and 21~nm, 51~nm, and 48~nm (Sample III, right panel, top left, bottom left, and top right QDs, respectively). (d) Photoluminescence spectra collected from the QDs shown in panel (a), collected under 780~nm continuous-wave laser excitation at a temperature of 4~K, on a silicon CCD camera.}
\label{fig3}
\end{figure*}

\begin{table*}[t]

\caption{Summary table showing the different samples under study, the number of QDs that have been analyzed (photoluminescence positioning, spectral characterisation, and atomic force microscopy), the type of surface features observed with the atomic force microscope (examples of "oval, "dip" and "crater" features are shown in Fig.\,2d, 3c (first column), 3c (second column) respectively), and the number of instances in which the stated AFM surface feature type was observed in proximity to the QD's emitting dipole.}
\centering
\vspace{0.1cm}
\begin{tabular}{|c|c|c|c|}

\hline
\vspace*{\fill} \textbf{Sample} &   \textbf{No. of QDs analyzed} &  \vspace*{\fill}  \textbf{AFM feature type}  &    \textbf{No. of matching AFM/QD positions} \\
 \hline
  I   & 25 & oval  & 7\\
 \hline
  II & 21 & dip    &5\\
 \hline
 III   & 33    & crater    &19\\
 \hline
  IV & 10 & oval    &6\\
 \hline
 \end{tabular}
\end{table*}

In Ref.\,\onlinecite{QD_oval}, the authors attribute the improved emission properties of the QD under study to its self-alignment to the center of a surface oval defect. By using our combined optical positioning-AFM approach, we are able to locate the QD emitting dipole with respect to the surface topography. For the oval defect shown in Fig.\,2(c) and 2(d), we see that the emitter is located at the periphery of the oval defect. By carrying out photoluminescence measurements on the QD, under laser excitation, we do not observe a higher brightness compared to other emitters on the same sample, possibly due to the misalignment of the QD with respect to the center of the oval defect that we observe.

When measuring two other samples (labeled II and III), we see different surface features in correspondence to QD locations. It is worth noting that each kind of feature is peculiar to a specific wafer growth. Figure\,3 shows examples of photoluminescence and AFM images, as well as the corresponding photoluminescence spectra collected from the specific QDs under study. The photoluminescence spectra are all characterized by sharp emission lines, as expected for high quality single QDs. In the first column (sample II), an example of a shallow and sharp dip in the AFM image is observed: by overlaying the position obtained from the photoluminescence image in panel (a) to the AFM image, we can see that the QD is aligned with the lowest portion of the dip. The second and third columns show surface features observed in a different sample (III), characterized by larger and more circular crater-like features. As shown in the third column, when larger AFM features are observed, several single QDs or clusters are likely to be seen. The presence of several optically active QDs is also confirmed by the photoluminescence spectra, where three groups of sharp peaks are visible (see Fig.\,3(d)).

\begin{figure*}[t]
 \includegraphics[width=0.6\linewidth]{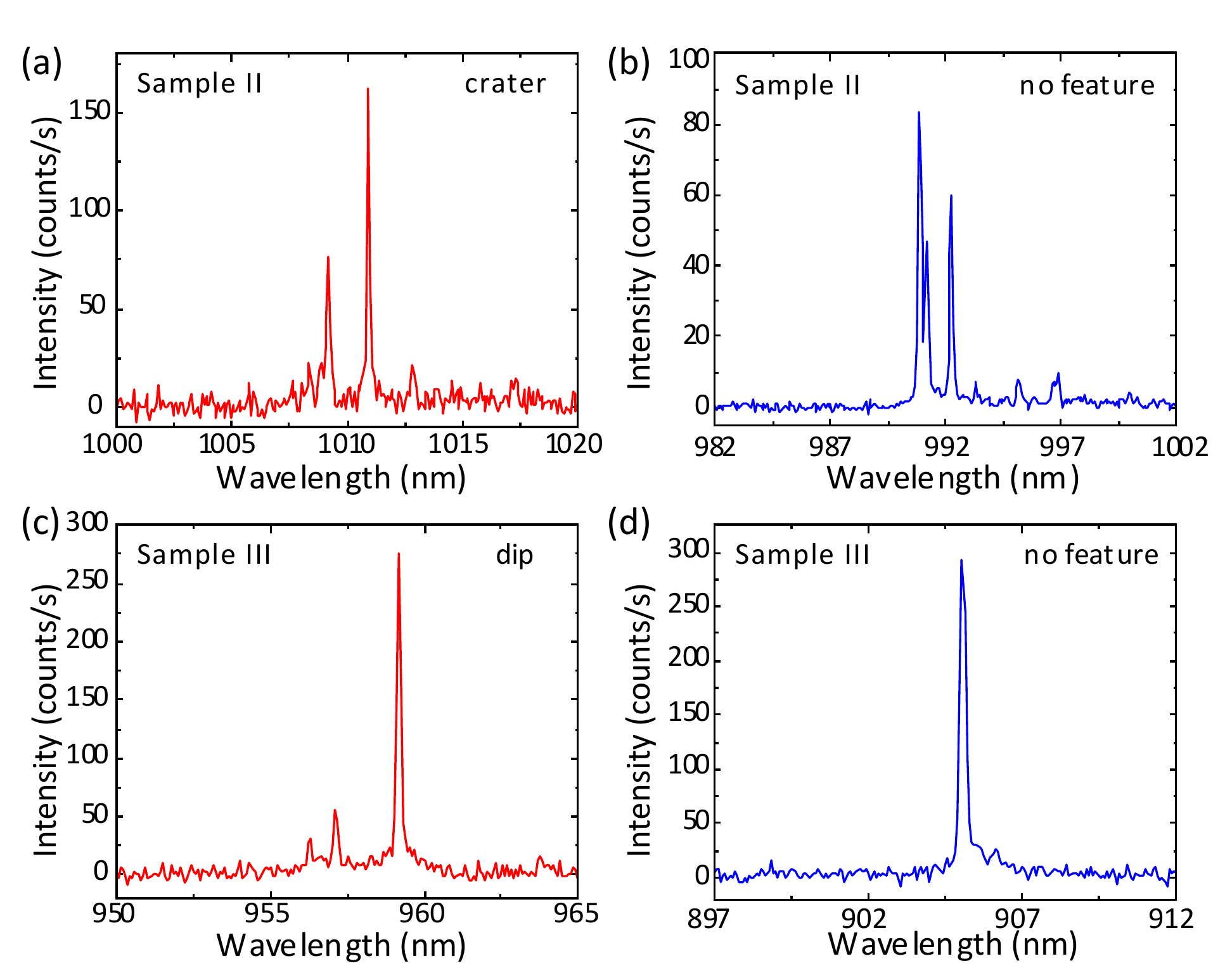}
   \caption{(a) and (c): Photoluminescence spectra, collected under non-resonant laser excitation at a temperature of 4~K, from single QDs aligned with surface features [dip-like (Sample II) in (a) and crater-like (Sample III) in (c)]. (b) and (d): Photoluminescence spectra from single QDs, collected from the same samples as the corresponding spectra in (a) and (c), but from areas not showing any surface features.}
   \label{fig4}
\end{figure*}

Similar topographical features are repeatedly observed for the same sample growth (oval defects in sample I, sharp shallow dips in sample II and larger crater-like features in sample III). In all cases, the sizes of the AFM features are too large to be caused solely by the presence of the buried QDs.  Instead, it seems likely that the features are indicative of other defects that occur during the MBE growth process.  This could include oval defect protrusions (sample I) that result from Ga droplets, for example, or various types of indentations that have been observed in GaAs MBE growth~\cite{surf_defects}. Such defects, if created prior to the InAs deposition step, might propagate upwards and seed QD nucleation, in a manner analogous to that which has been deterministically exploited in the growth of site-controlled QDs on patterned surfaces \cite{Kapon, Kapon2}.  This could then explain the correlation between QD emission and topographic features in the AFM images.  It is worth noting that the QD containing layer is grown on top of a thick AlGaAs layer (in place for subsequent undercut processes when fabricating devices), and that the growth of high Al-content AlGaAs layers can be one source of defects~\cite{AlGaAs_defects}.

In Table I, the results of our study are summarized: a significant fraction of the examined locations, ranging from about 20\,$\%$ to 60\,$\%$, shows surface topographical features that correspond with the QD locations, thus implying that such topographical defects are not solely responsible for QD nucleation but in many cases might behave as catalysts to the growth process. These surface features are large enough to potentially influence nanophotonic geometries that might be subsequently fabricated to exercise control on the QD emission pattern and lifetime (e.g., microresonators and photonic crystals). However, they have no apparent influence on the basic QD emission properties under non-resonant excitation. Photoluminescence spectra, collected from single emitters in areas where detectable surface features are both visible and absent, are shown in Fig.\,4. In both cases, the linewidths are resolution-limited and no specific difference is observed. Given that, from photoluminescence measurements, it is not possible to determine which emitter is aligned with potentially large and pronounced surface features, we anticipate that our approach, combining optical and AFM characterization, may be valuable in selecting single QDs to be embedded within optimized nanophotonic devices for quantum information technology applications.

\section{Discussion}

The importance of the surface topology when fabricating photonic devices for quantum information technology applications is multifold. Coherence properties can be affected by defect states that might appear in correspondence with pronounced surface features \cite{Jin}. The presence of charged defects can affect the coherence properties of the single-photon emission and even induce blinking \cite{Marcelo_blink} and spectral diffusion \cite{Richard}.
The presence of surface topographical defects of sizes that can be larger than 1\,$\mu$m may affect the far-field emission pattern in a beneficial way \cite{QD_oval} but may also be detrimental if the emitter is not aligned with the center of the oval defect or is aligned with crater-like defects. This is particularly relevant when dealing with nanoscale devices, such as photonic crystals or dielectric gratings, where the device features can be comparable to the length scale of the surface defects.  To verify this, we carry out finite-difference time-domain simulations of circular grating 'bullseye' microcavities (such as those in Ref.~\onlinecite{NComm}) in which surface defects of an idealized shape (semi-ellipsoidal craters) and similar length scale as those observed in our experiments are introduced. The results indicate a shift in resonant wavelength of up to 20~nm, a reduction of the fraction of emission that is in the upwards vertical direction, and a modification of the far-field emission pattern in comparison to unperturbed devices. Furthermore, if the QD nucleates in a dip in the GaAs substrate, the emitter's vertical position can be lower than the nominal one. This is a factor that should be taken into account in the electromagnetic simulations to obtain optimal performance, if a photonic device is to be fabricated around that specific QD.

Our combined photoluminescence-AFM technique can be applied to select suitable QD emitters to be integrated into photonic devices, by pre-screening the surface topography. This is particularly relevant in cases when more advanced optical techniques, such as scanning Fabry-Perot interferometry that would provide information about the coherence of the emission (which may be reduced by the presence of defects), can be difficult to implement, given the low intensity of the signal emitted by nanostructures in bulk. Such an approach is therefore expected to be beneficial for increasing the yield of fabricated photonic devices with optimal performances.
Once the QD is located, more advanced AFM techniques, such as multifrequency AFM to study subsurface properties \cite{Santiago} or contact mode force measurements \cite{AFM_force} to assess indentation levels and therefore strain properties, could also be implemented to extract structural information on the properties of buried QD samples.

\vspace{0.1in}

\section*{Acknowledgements}
We would like to thank Marcelo Davan\c co for useful discussions, Santiago D. Solares and Tobias Meier for advice on AFM techniques, and Ralf Schniersmeier for technical support.  LS acknowledges financial support from EPSRC, grant EP/P001343/1. JL acknowledges support under the Cooperative Research Agreement between the University of Maryland and NIST-CNST, Award 70NANB10H193. JDS acknowledges support from the KIST flagship institutional program.

 \end{document}